\documentclass[pre,twocolumn,aps,eqsecnum]{revtex4}
\usepackage{epsfig}

\begin{document}

\title{Anomalous diffusion and stretched exponentials in heterogeneous glass-forming liquids: Low-temperature behavior}

\author{ J. S. Langer and Swagatam Mukhopadhyay}
\affiliation{Dept. of Physics, University of California, Santa
Barbara, CA  93106-9530}

\date{\today}

\begin{abstract}
We propose a model of a heterogeneous glass forming liquid and compute the low-temperature behavior of a tagged molecule moving within it.  This model exhibits stretched-exponential decay of the wavenumber-dependent, self intermediate scattering function in the limit of long times.  At temperatures close to the glass transition, where the heterogeneities are much larger in extent than the molecular spacing, the time dependence of the scattering function crosses over from stretched-exponential decay with an index $b=1/2$ at large wave numbers to normal, diffusive behavior with $b = 1$ at small wavenumbers.  There is a clear separation between early-stage, cage-breaking $\beta$ relaxation and late-stage $\alpha$ relaxation. The spatial representation of the scattering function exhibits an anomalously broad exponential (non-Gaussian) tail for sufficiently large values of the molecular displacement at all finite times.  

\end{abstract}

\maketitle

\section{Introduction}
\label{intro}
A growing body of evidence, both experimental \cite{Ediger-ARPC-00,BERTHIER05,Weeks-Weitz-Science-00} and numerical \cite{Kob-Glotzer-PRL-97,Yamamoto-Onuki-PRE-98}, points to the existence of intrinsic spatial heterogeneities in glass forming liquids at temperatures slightly above the glass transition, and the relevance of these heterogeneities to stretched exponential decay of correlations and anomalous, non-Gaussian diffusion.  Here we propose a simple model of a heterogeneous glass former and use it to compute observable properties of a tagged molecule moving within it.  In particular, we show that this model naturally predicts stretched-exponential decay when the scale of the heterogeneity is much larger than the molecular spacing; and we compute deviations from Gaussian displacement distributions during intermediate stages of the relaxation process.

Our model is loosely motivated by the excitation-chain (XC) theory of the glass transition proposed recently by one of us (JSL) \cite{Langer-PRE-06,Langer-PRL-06}  This theory suggests that a glass-forming liquid at a temperature $T$ not too far above the glass transition temperature $T_0$ consists of fluctuating domains of linear size $R^*(T)$, within which the molecules are  frozen in a glassy state where they have little or no mobility.   $R^*(T)$ is a length scale that characterizes dynamic fluctuations in the equilibrium states of these systems.   The theory predicts that $R^*(T)$  diverges near $T_0$ like $(T-T_0)^{-1}$ and decreases to zero at the upper limit of the super-Arrhenius region, $T=T_A$. 

Characterizing the more mobile material that lies outside the hypothetical frozen domains remains one of the deeper challenges in glass physics.  In the language introduced by Widmer-Cooper and Harrowell \cite{HARROWELL}, the mobile molecules reside in regions of high ``propensity.''  Those authors perform molecular dynamics simulations of a glasslike, two-dimensional, binary mixture.  They average the mean-square displacement of each molecule over an isoconfigurational ensemble in which the initial positions of all the molecules are fixed, but the initial velocities are selected at random from a Boltzmann distribution.  The propensity of a molecule is its mean-square displacement during a structural relaxation time.  Contour maps of propensity as a function of initial molecular positions do indeed exhibit well defined domains of low and high mobility.   

Widmer-Cooper and Harrowell compare their propensity maps with corresponding plots of the potential energy, which is taken to be indicative of the local structure. Interestingly, the  propensity appears to be almost entirely uncorrelated with the potential energy.  In retrospect, this lack of correlation between short-range structure and mobility may not be surprising.  Such a correlation would be expected for a slowly coarsening system consisting of domains of two structurally distinct phases, for example, a slightly supersaturated (mobile) fluid in which (frozen) crystallites are growing.  In that case, the domains are very nearly but not quite in thermodynamic coexistence with each other.  A true glass-forming liquid,  however, is in a state of thermodynamic equilibrium.  The temperatures and chemical potentials are uniform everywhere, independent of propensity; and the local pair correlations, which ought to be determined primarily by the temperature in an equilibrium state, appear to be statistically uniform as well.  

A local quantity that does correlate with the propensity is the mean-square vibration amplitude of a molecule averaged over times much longer than its period of oscillation but much shorter than the time required for irreversible rearrangements to occur.  A large mean-square vibration amplitude -- or, equivalently, a small Debye-Waller factor -- implies that a molecule is participating in a soft, low frequency, elastic mode.  It is well known that soft modes are abundant in amorphous materials near jamming transitions.\cite{SILBERT} Moreover, it appears (for example, in Fig.4 of \cite{SILBERT}) that the largest displacements in such modes lie predominantly along one dimensional, chainlike paths.  Thus, regions of high propensity may be elastically soft, and may somehow be correlated with excitation-chain activity.  Of course, local variations of elastic stiffness are ultimately structural in nature; but soft modes are collective phenomena involving many molecules, and are not easily detected by measuring near-neighbor pair correlations.  

In the model proposed here, glassy domains of low propensity are separated from each other by regions of higher propensity in which many of the molecules are mobile.  This domain structure undergoes persistent fluctuations on time scales of the order of $\tau_{\alpha}^* = {R^*}^2/D_{\alpha}$, where $D_{\alpha}= \ell^2/\tau_{\alpha}$ is the diffusion constant associated with $\alpha$ relaxation, $\ell$ is approximately the average molecular spacing, and $\tau_{\alpha}$ is the strongly temperature-dependent, super-Arrhenius, $\alpha$ relaxation time. In contrast to the slow fluctuations of domain boundaries,  individual molecules within high-propensity regions are random walkers  with a diffusion constant $D_M =\ell^2/\tau_M$,  where $\tau_M$ is a temperature-dependent time scale -- presumably not super-Arrhenius -- that characterizes mobile molecular displacements. 

To a first approximation, a tagged molecule at any given time is either frozen in a glassy domain or is mobile.  If the former, then it waits a time of the order of $\tau_{\alpha}^*$ before it is encountered by mobile neighbors and itself starts to undergo displacements.  If the latter, then it diffuses for some distance until it finds itself frozen again in a glassy domain. That distance is at least a few molecular spacings $\ell$; but it seems more  likely to be proportional to $R^*$ if the domains are large and if diffusive hopping takes place more readily within the mobile regions than across their boundaries and into the glassy domains.  Therefore, we assume that the time elapsed during mobile motion is of the order of ${R^*}^2/D_M = (R^*/\ell)^2\,\tau_M$.  The ratio of these two time scales is $\tau_{\alpha}/\tau_M   \equiv \Delta$. At temperatures well below $T_A$, we expect that $\Delta \gg 1$.  

A similar picture of diffusion in fluctuating regions of varying mobility emerges in  kinetically constrained models \cite{JUNG-GARRAHAN-CHANDLER-04,BERTHIER-GARRAHAN-CHANDLER-05} and in a recent model of a gel \cite{HURTADO-BERTHIER-KOB-07}.  Our analysis is most closely related to the diffusion model of Chaudhuri et al \cite{CHAUDHURIetal07}, which has been a valuable  starting point for the present investigation.  We go beyond \cite{CHAUDHURIetal07} by including a mechanism for producing late-stage stretched-exponential relaxation, and by distinguishing that mechanism from the diffusion of mobile molecules.  In contrast to \cite{CHAUDHURIetal07}, our molecules switch back and forth between frozen and mobile behavior and thus explore the geometry of the domain structure.  As a result, our model exhibits a clear separation between the slow $\alpha$ relaxation and the faster mobile motions that we interpret as $\beta$ relaxation. It also respects time-translational symmetry, which seems to be violated in \cite{CHAUDHURIetal07}.  

The preceding discussion pertains just to the case of large-scale heterogeneities at temperatures only a little above $T_0$ and well below $T_A$.  Some of the most interesting and experimentally accessible physics, however, occurs near the crossover to small-scale heterogeneity near $T_A$.  Constructing a theory of diffusion in the crossover region requires a  temperature-dependent analysis, which will be described in a following paper.

\section{Basic Ingredients of the Model in the Limit of Large-Scale Heterogeneity}
\label{basicmodel}

As outlined above, there are two different mechanisms to be described probabilistically in this sytem, each operating on its own characteristic time scale.  The separation between time scales, characterized by a large value of $\Delta$,  suggests that the natural mathematical language in which to discuss this system is that of a continuous time random walk (CTRW) in which a tagged molecule alternates between long waits in glassy domains and faster displacements in mobile regions.  In this Section, we compute the waiting time distributions for molecules in glassy and mobile domains, and the corresponding probability distributions for the mobile displacements.  

We immediately encounter a problem, however, because there are more than just two characteristic time scales in this system.  Analog experiments \cite {Weeks-Weitz-Science-00} and numerical simulations \cite{Kob-Glotzer-PRL-97} typically follow the motions of particles starting from their initial positions, and resolve these trajectories on length scales smaller than the interparticle spacing, and on time scales of the order of the shortest vibrational periods.  This is a continuous range of scales.  At its long-time limit, it includes the slow modes that are involved in molecular rearrangements, whose amplitudes may exceed the Lindemann melting criterion in the mobile regions.  Thus, the cage-breaking, $\beta$-relaxation mechanism may be part of the same continuous range of time scales that includes the intra-cage vibrational motions.  Indeed, no sharp distinction between vibrational and cage breaking time scales seems to appear in experimental data. If these are not clearly distinct kinds of events, then the CTRW approximation is not valid in the short-time limit.  

In view of this difficulty, we choose here to consider only the $\beta$ and $\alpha$ time scales, $\tau_M$ and $\tau_{\alpha}$ respectively.  By neglecting the shorter time scales, i.e. those of the order of $\tau_0 \sim$ femtoseconds - picoseconds, we imply that we are not resolving length scales much smaller than the molecular spacing $\ell$.  We then assume that the probability that a molecule has left its cage at times of order $\tau_M \gg \tau_0$ is determined by the probability that it has mobile neighbors and participates in local molecular rearrangements.  At the longest time scales $\tau_{\alpha} \sim$ seconds (by definition, the viscous relaxation time at the glass temperature), the probability that a molecule has escaped from a large glassy domain is determined by the probability that it has been encountered by the boundary of the domain.  

Consider first the slowest motions, i.e. those associated with fluctuations of the domain boundaries on length scales $R^*$ and time scales $\tau_{\alpha}^*$.  Define $t^*\equiv t/\tau_{\alpha}^*$, where $t$ is the physical time in seconds; and let $\psi_G(t^*)$ be the normalized probability distribution for the time that a molecule spends in a glassy domain before entering a mobile region.  Write this distribution in the form
\begin{equation}
\label{psiG} 
\psi_G(t^*) = \int_0^{\infty} d\rho\,W(\rho)\,{e^{-t^*/\rho^2}\over \rho^2},
\end{equation}
where $\rho$ is the linear size of a domain in units $R^*$. $W(\rho)$ is a normalized distribution over these sizes, and the remaining factor inside the integrand is a normalized distribution over $t^*$. The quantity $\rho^{-2}$ appearing in the exponential in Eq.(\ref{psiG}) is the lowest eigenvalue of the diffusion kernel for a molecule moving in a domain of size $\rho$.  Note the similarity to the trapping models discussed, for example, in \cite{G-P,Monthus-Bouchaud-96,BENDLER02}; but also note that the ``trap'' here is a two dimensional subspace bounding a three dimensional domain. The quantity that is fluctuating in a normal diffusive manner is the mean square of the distance  between the tagged molecule and the domain boundary.  Because the boundary surrounds the molecule, the direction of the diffusive drift is irrelevant; any point of contact on the boundary is equivalent to any other.  Therefore, it makes little difference whether the molecule is the diffuser and the boundary is the target, or -- as in this case -- the boundary diffuses and the molecule is the target. 

Strictly speaking, Eq.(\ref{psiG}) is a long-time approximation; at shorter times, the higher eigenmodes of the diffusion kernel make non-negligible contributions.  However, this approximation is qualitatively adequate for our purposes at shorter times as well because $\psi_G(t^*)$ is well behaved at small $t^*$ and, as mentioned previously, we are not trying to include the short-time behavior in this function.  If we assume that the distribution over values of $\rho$ is Gaussian, $W(\rho) \propto \rho^2\,\exp(-\rho^2)$ in three dimensions, then
\begin{equation}
\label{psiG2}
\psi_G(t^*)= 2\,e^{-2\,\sqrt{t^*}}.
\end{equation}
That is, we find a rudimentary but nontrivial stretched exponential of the form $\exp\,(- {\rm const.}\times {t^*}^b)$ with $b = 1/2$. 

There is little  reason to expect that $W(\rho)$ remains Gaussian out to the large values of $\rho$ that determine the long-time behavior of $\psi_G(t^*)$.  Deviations from the Gaussian would produce different indices.  For example, as the temperature increases and the glassy domains become smaller, their size distributions might cut off more sharply than a Gaussian.  Thus, if $W(\rho) \propto \exp(-\rho^m)$, with $m \ge 2$, then $b = m/(m+2) \to  1$ as $m \to \infty$.  The problem of computing this distribution or, equivalently, values of $m$ from first principles may eventually become solvable as we learn more about the statistical mechanics of glass forming liquids; but that problem is beyond the scope of this investigation.  For present purposes, the important points are that a plausible distribution $W(\rho)$ produces stretched-exponential behavior, and that the mechanism by which this happens could produce a temperature dependent index $b$.

Next, consider the waiting-time distribution $\psi_M(t^*)$ for a molecule in a mobile region.  This situation is qualitatively different from that of a molecule in a glassy domain because the distance traveled by the molecule is related to the time during which it remains mobile. As stated earlier, we assume that a mobile molecule diffuses a distance of order $R^*$ before reentering a glassy region, so that its residence time in the mobile region is of order $\tau_M\,(R^*/\ell)^2$ which, in $t^*$ units, is simply $\Delta^{-1}$.  Then, for simplicity, assume an exponential waiting-time distribution:
\begin{equation}
\label{psiM}
\psi_M(t^*) = \Delta\,e^{-\Delta\,t^*}.
\end{equation}
Compared to $\psi_G(t^*)$ in Eq.(\ref{psiG2}), $\psi_M(t^*)$ is sharply peaked near $t^*=0$ if, as expected, $\Delta$ is large.  To complete the model of mobile motion, we need the conditional probability $p_M(r^*,t^*)$ for diffusion over a scaled distance $r^*=r/R^*$ in time $t^*$.  The natural choice is
\begin{equation}
\label{pM}
p_M(r^*,t^*)= {1\over (2\,\pi\,\Delta\,t^*)^{3/2}}\,\exp\,\left(- {{r^*}^2\over 2\,\Delta\,t^*}\right),
\end{equation}
which is a normalized, three dimensional distribution over $r^*$.

\section{Continuous-Time Random Walks}
\label{CTRW}

The next step is to translate these physical ingredients of the model into the language of  continuous-time random walks.\cite{MONTROLL-SHLESINGER-84,BOUCHAUD-GEORGES-90}  Define two different probability distribution functions, $n_G(r^*,t^*)$ and $n_M(r^*,t^*)$, for molecules starting, respectively, in glassy domains or mobile regions, and moving distances $r^*$ in times $t^*$.  Each molecular trajectory consists of a sequence of transitions between glassy domains and mobile regions.  A single transition in which a molecule starts in a glassy domain at time $t_1^*$ and ends in a mobile region at time $t_2^*$ (without having changed its actual position) occurs with probability $\psi_G(t_2^*-t_1^*)$.   Similarly, a transition in which a molecule starts in a mobile region at time $t_1^* $ and position ${\bf r}_1^*$, and  ends in a glassy domain at time $t_2^*$ and position ${\bf r}_2^*$, has probability 
$\psi_M(t_2^*-t_1^*)\, p_M(|{\bf r}_2^*-{\bf r}_1^*|, t_2^*-t_1^*)$.  The probability that a molecule starts or arrives in a glassy domain at time $t_1^*$ and is still there at the final time $t^*$ is $\phi_G(t^*-t_1^*)$, where
\begin{equation}
\label{phiG}
\phi_G(t^*)=\int_{t^*}^{\infty} \psi_G({t^*}')\,d{t^*}'= (1+2\,\sqrt{t^*})\,e^{-2\,\sqrt{t^*}}.
\end{equation}
Finally, the probability that a molecule starts or arrives in a mobile region at time $t_1^*$ and position ${\bf r}_1^*$ and is still in that region at the final time $t^*$ and at position ${\bf r}^*$, is 
$\phi_M(t^*-t_1^*)\,p_M(|{\bf r}^*-{\bf r}_1^*|, t^*-t_1^*)$,
where, in analogy to Eq.(\ref{phiG}),
\begin{equation}
\label{phiM}
\phi_M(t^*)=\int_{t^*}^{\infty} \psi_M(t_1^*)\,d t_1^* = e^{-\Delta\,t^*}.
\end{equation}

The multiple convolutions that describe each trajectory are, as usual, converted into products by computing Fourier-Laplace transforms of each function:
\begin{equation}
\tilde n_j(k,u)=\int e^{-i{\bf k}\cdot{\bf r^*}}d{\bf r^*}\int_0^{\infty} e^{-ut^*} n_j(r^*,t^*)\,dt^*,
\end{equation}
where $j = G,M$.  For jumps starting in glassy regions, we need: 
\begin{equation}
\label{psiGu}
\tilde \psi_G(u)=\int_0^{\infty} e^{-ut^*}\psi_G(t^*)\,dt^*,
\end{equation}
and
\begin{equation}
\label{phiGu}
\tilde\phi_G(u)={1-\tilde\psi_G(u)\over u}.
\end{equation}
For the mobile regions, the spatial Fourier transform of the conditional probability distribution $p_M(r^*,t^*)$ is 
\begin{equation}
\hat p_M(k,t^*)=\exp\,\left(-{\Delta\,k^2\,t^*\over 2}\right);
\end{equation}
therefore, define
\begin{eqnarray}
\nonumber
f_M(k,u)&\equiv& \int_0^{\infty} e^{-ut^*} \hat p_M(k,t^*)\,\psi_M(t^*)\,dt^*\\ \cr&=&{\Delta\over u+\Delta\,(1+k^2/2)},
\end{eqnarray}
and
\begin{eqnarray}
\nonumber
g_M(k,u)&\equiv&\int_0^{\infty} e^{-ut^*} \hat p_M(k,t^*)\,\phi_M(t^*)\,dt\\ \cr&=&{1\over u+\Delta\,(1+k^2/2)}.
\end{eqnarray}

Putting these pieces together, and summing over indefinitely many individual jumps in each trajectory, we find
\begin{eqnarray}
\label{nGku}
\nonumber
\tilde n_G(k,u)&=& {\tilde \phi_G(u)+g_M(k,u)\,\tilde \psi_G(u)\over 1-\tilde \psi_G(u)\,f_M(k,u)}\\ \cr &=&{1\over u}\,{N_G(k,u)\over W(k,u)};
\end{eqnarray}
and
\begin{eqnarray}
\label{nMku}
\nonumber
\tilde n_M(k,u)&=& {\tilde \phi_G(u)\,f_M(k,u)+ g_M(k,u)\over 1-\tilde \psi_G(u)\,f_M(k,u)}\\ \cr &=&{1\over u}\,{N_M(k,u)\over W(k,u)};~~~~
\end{eqnarray}
where
\begin{equation}
N_G(k,u) = \bigl[1-\tilde\psi_G(u)\bigr](1+k^2/2) + u/\Delta;
\end{equation}
\begin{equation}
N_M(k,u)= 1-\tilde\psi_G(u)+u/\Delta;
\end{equation}
and
\begin{equation}
W(k,u)= 1-\tilde\psi_G(u)\,+k^2/2+u/\Delta. 
\end{equation}
In both Eqs.(\ref{nGku}) and (\ref{nMku}), in the numerators of the first expressions, the first term corresponds to trajectories that end in glassy domains, and the second to those that end in mobile regions.  

\begin{figure}[h]
\centering \includegraphics[height=7 cm]{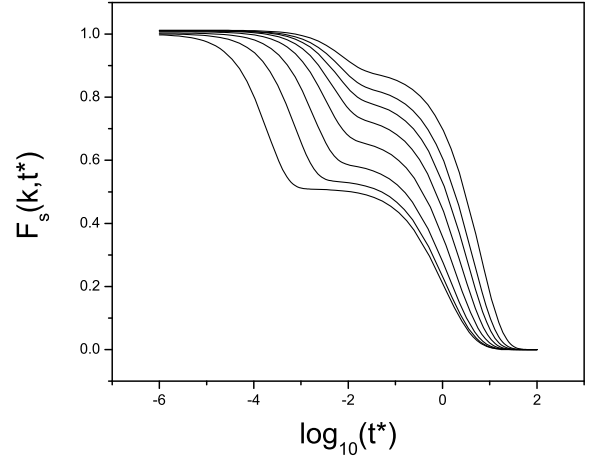}
\caption{\label{ISFfig} Intermediate scattering function $F_s(k,t^*)$ with ${\cal P}_G = 0.5$ and $\Delta = 100$, for $k = 0.8,\,1,\,1.2,\,1.5,\,2,\,3,\,5,\,10$, reading from top to bottom.}
\end{figure}

\section{Intermediate Scattering Function}
\label{ISFsec}

The conventional way to study the diffusion mechanisms discussed here is to measure the self intermediate scattering function $F_s(k,t)$, which, in the present notation, is the mixed Fourier-time representation of the weighted average of $n_G$ and $n_M$.  That is,
\begin{equation}
\label{ISFdef}
F_s(k,t^*)= {\cal P}_G\,\hat n_G(k,t^*)+ (1-{\cal P}_G)\,\hat n_M(k,t^*),
\end{equation}
where ${\cal P}_G$ is the probability that a molecule starts its motion in a frozen, glassy environment; and
\begin{equation}
\label{hatndef}
\hat n_j(k,t^*)= \int_{- i\infty}^{+i\infty}{du\over 2\pi i}\,e^{u\,t^*}\,\tilde n_j(k,u),~~~j = G,\,M.
\end{equation} 

The inversions of Laplace-transforms in Eq.(\ref{hatndef}) are nontrivial because the stretched-exponential function $\tilde\psi_G(u)$, defined in Eq.(\ref{psiGu}), has an essential singularity at $u=0$.  For $t^* > 0$, the integrations over $u$ in Eq.(\ref{hatndef}) must be performed by closing the contour in the negative half $u$-plane, which requires that $\tilde n_j(k,u)$ be analytically continued to points $u = |u|\,\exp(i\theta)$. To evaluate $\tilde\psi_G(u)$ at such points, using Eq.(\ref{psiGu}), let $t^*=y^2$ and then rotate the $y$ contour away from the positive real axis to a line $y=\xi\,\exp(-i\theta/2)$, $0<\xi<\infty$ so that the integrand always decreases rapidly at infinity. For example, for $u\to -w\pm i\epsilon$, $\theta = \pm \pi$, 

\begin{eqnarray}
\label{A+iB}
\nonumber
&&\tilde \psi_G(-w\pm i\epsilon)=-4\int_0^{\infty}\xi\,d\xi\,e^{-w \xi^2\pm 2 i \xi}\cr &&= 
-{2\over w}+{2\sqrt{\pi}\over w^{3/2}}\,e^{-1/w}\left({\rm Erfi}\Bigl({1\over\sqrt{w}}\Bigr)
\mp \,i\right)\\ \cr&&\equiv A(w) \mp i\,B(w),
\end{eqnarray}
where Erfi is the imaginary error function.  The imaginary part of $\tilde \psi_G(-w\pm i\epsilon)$ is the discontinuity across a cut along the negative $u$ axis. With this formula, it is straightforward to compute the discontinuity across the cut for the $\tilde n_j(k,u)$ in Eqs.(\ref{nGku}) and (\ref{nMku}), and in this way to compute the scattering functions by closing the contour around this cut. That is:
\begin{eqnarray}
\label{hatnj}
\nonumber
\hat n_j(k,t^*)&=& \int_0^{\infty}dw\,{e^{-w\,t^*}\over \pi\,w}\\ \cr &\times&{\rm Im} \left[{N_j(k,-w+i0)\over W(k,-w+i0)}\right],~~j=G,\,M.~~~~~~~
\end{eqnarray}
The integrands are well behaved as $w \to 0^+$, therefore it is possible to let the lower limit of integration be $w=0$.  

The graphs in Fig.\ref{ISFfig} show $F_s(k,t^*)$, computed numerically from Eq.(\ref{hatnj}), as a function of $\log_{10}(t^*)$, for $\Delta = 100$, ${\cal P}_G = 0.5$, and for a sequence of wavenumbers $k$. These are low-temperature scattering functions, where $\Delta \gg 1$, which means that molecules spend much longer times frozen in glassy domains than they do when moving in mobile regions.  The  choice ${\cal P}_G=0.5$ is made primarily for clarity; but it seems likely that this value of ${\cal P}_G$ describes a system that is well into its non-Arrhenius, anomalous-diffusion regime.  The set of curves in Fig.\ref{ISFfig} closely corresponds to those shown, for example, in \cite{KA}.  Howewver, as explained in Sec.\ref{basicmodel}, we do not claim to resolve the small-time behavior accurately.  

For the larger values of $k$ shown in Fig.\ref{ISFfig}, the scattering functions exhibit two-stage relaxation. The first drop, which we identify as the $\beta$ relaxation, occurs because the initially mobile molecules are diffusing beyond their cages. Then, after much longer times indicated by the plateaus, the scattering curves cross over to typical $\alpha$ behavior in which all molecules, independent of whether or not they were initially mobile, make slow transitions back and forth between mobile and frozen states.  This two-step relaxation disappears at small $k$, where the scattering functions are averages over distances larger than $R^*$ and exhibit normal, diffusive behavior.  

\begin{figure}[h]
\centering \includegraphics[height=7 cm]{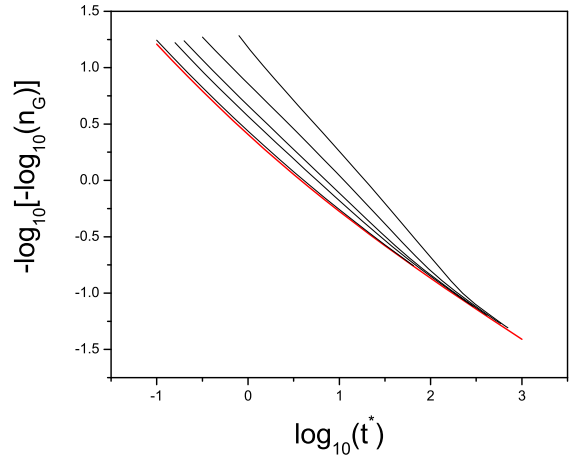}
\caption{\label{xG} (Color online) Intermediate scattering function $\hat n_G(k,t^*)$ for $k = 0.7,\,1,\,1.5,\,5,\, 10$, reading from top to bottom.  The red curve is the glassy waiting time distribution $\phi_G(t^*)$, which in this plot is indistinguishable from the large-$k$ limit of $\hat n_G(k,t^*)$.}
\end{figure}

The most interesting feature of these scattering functions is that they exhibit a continuous range of stretched exponential relaxation modes.  To see this, we next look at the long-time behavior of $\hat n_G(k,t^*)$ for relatively large values of $t^*$ in the range  $1 < t^* < 1000$, and for a sequence of values of $k$.  In Fig.\ref{xG}, we plot  $-\log_{10}\,[-\log_{10} \hat n_G(k,t^*)]$ as a function of $\log_{10}({t^*})$, so that the slopes of the curves are equal to (minus) the stretched-exponential index $b$. The results are again for $\Delta = 100$.

Each of the curves in Fig.\ref{xG} has a constant slope over about two decades in $t^*$, indicating apparently well defined values of $b$ in $\hat n_G\propto \exp\,(-{t^*}^b)$.  In the limit of large $k$, $\hat n_G(k,t^*)$ is indistinguishable from the glassy waiting time distribution $\phi_G(t^*)$ defined in Eq.(\ref{phiG}) and shown by the red curve in the figure; and $b$ is accurately equal to $1/2$ for large $t^*$.  $b$ increases toward unity as $k$ decreases toward values of order unity or smaller.  For any nonzero $k$, these curves also exhibit a crossover from $b > 1/2$ at small $t^*$ to $b = 1/2$ at large $t^*$. 

To see this behavior analytically, we can deduce from Eq.(\ref{A+iB}) that $A(w)\approx 1 + 3\,w/2$ for the very small values of $w$ that are relevant at very large $t^*$. Therefore, for large $\Delta$,
\begin{equation}
W(k,-w+i0)\approx {k^2\over 2} - {3\,w\over 2} + i\,B(w);
\end{equation}
and, because $B(w)\ll 1$ for $w \ll 1$, the integrand in Eq.(\ref{hatnj}) is sharply peaked at $w = k^2/3\,$  so long as $k^2 \ll 1$.  For $t^*\gg 1$, the integrand in Eq.(\ref{hatnj}) has another sharp peak at the maximum of the function $\exp\,(- w\,t^*)\,B(w)$, i.e. at $w = w^*= 1/\sqrt{t^*}$.  If $k^2/3 < 1/\sqrt{t^*} \ll 1$, then the diffusive peak is dominant, and $\hat n_G \sim \exp\,(- k^2\,t^*/3)$.  On the other hand, if $ 1/\sqrt{t^*} < k^2/3 \ll 1$, then the anomalous peak at $w^*$ is dominant, and a saddle-point estimate yields $\hat n_G \sim 2\,\sqrt{t^*}\,\exp\,(- 2\,\sqrt{t^*})$ as expected from Eq.(\ref{phiG}).  Thus, at any fixed large time $t^*$, $\hat n_G$ becomes purely diffusive in the limit of small $k$.  Conversely, at fixed small $k$, the diffusion becomes anomalous at large enough $t^*$. 
\begin{figure}[h]
\centering \includegraphics[height=7 cm]{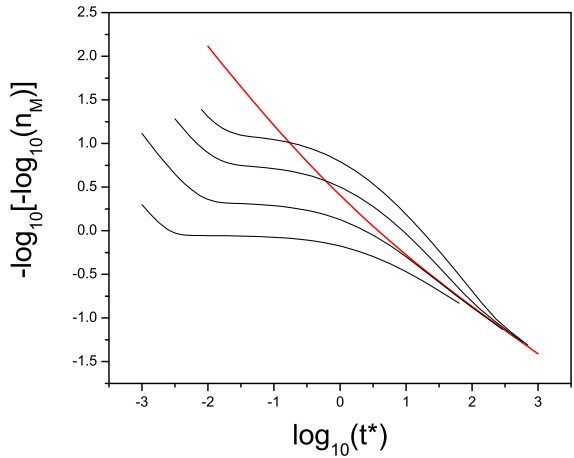}
\caption{\label{xM} (Color online) Intermediate scattering function $\hat n_M(k,t^*)$ for $k = 0.7,\,1,\,2,\,5$, reading from top to bottom.  The red curve is the glassy waiting time distribution $\phi_G(t^*)$.}
\end{figure}

A complementary set of behaviors is illustrated in Fig.\ref{xM}, which is the analog of Fig.\ref{xG} for the initially mobile molecules.  That is, Fig.\ref{xM} shows $-\log_{10}\,[-\log_{10} \hat n_M(k,t^*)]$ as a function of $\log_{10}({t^*})$, but for a broader range of times extending in two cases down to $t^* \sim 10^{-3}$.  At large times, the preceding analysis remains valid; for large $k$, $\hat n_M \sim \phi_G(t^*)\sim 2\,\sqrt{t^*}\,\exp\,(-2\,\sqrt{t^*})$.  As may be expected, however,  and consistent with the behavior seen in Fig.\ref{ISFfig}, $\hat n_M$ decays diffusively at small times (with slope $-1$ in this graph), and then enters a flat plateau before crossing over to the long-time anomalous behavior characteristic of both of these scattering functions.  For large enough $k$ and/or $\Delta$, the early-time  diffusive behavior can be deduced analytically.  In this case, the relevant values of $w$ are large, and $A(w) \approx -2/w$ becomes negligibly small.  Therefore,
\begin{equation}
W(k,-w+i0)\approx  1 + {k^2\over 2} - {w\over\Delta} + i\,B(w),
\end{equation}
where, for large $w$, $B(w)\approx 2\,\sqrt{\pi}/w^{3/2}$ again becomes small. Now the integrand has a sharp peak at $w = \Delta\,(1 + k^2/2)$.  This peak dominates the integrand at large $w$ for $\hat n_M$, but its amplitude is smaller by a factor $\Delta^{-1}$ for $\hat n_G$.  Thus, for small $t^*$, $\hat n_M \sim \exp\,[- (1+k^2/2)\,\Delta\,t^*]$ is normally diffusive, while no such behavior occurs in $\hat n_G$.

\section{Spatial Distribution Functions}
\label{SDF}
Yet another view of the normal and anomalous diffusive behaviors is obtained by looking at the spatial distribution functions themselves, that is, by computing $n_G(r^*,t^*)$ and $n_M(r^*,t^*)$.  The Fourier transforms in Eqs.(\ref{nGku}) and (\ref{nMku}) can be inverted analytically, yielding
\begin{equation}
\label{nGrt}
n_G(r^*,t^*)=\phi_G(t^*)\,\delta({\bf r}^*)- {1\over 4\,\pi\,r^*}\,{\partial\over\partial r^*}\, \Gamma_G(r^*,t^*),
\end{equation}
where 
\begin{equation}
\label{GammaG}
\Gamma_G(r^*,t^*)=\int_{-i\infty}^{+i\infty}{du\over 2\pi\,i u}\,e^{u t^*}\,\tilde\psi_G(u)\,\kappa(u)\,e^{-\kappa(u)\,r^*},
\end{equation}
and
\begin{equation}
\kappa(u)=\sqrt{2\,\Bigl[1-\tilde\psi_G(u)+u/\Delta\Bigr]},~~~{\rm Re}\,\kappa\ge 0.
\end{equation}

\begin{figure}[h]
\centering \includegraphics[height=7 cm]{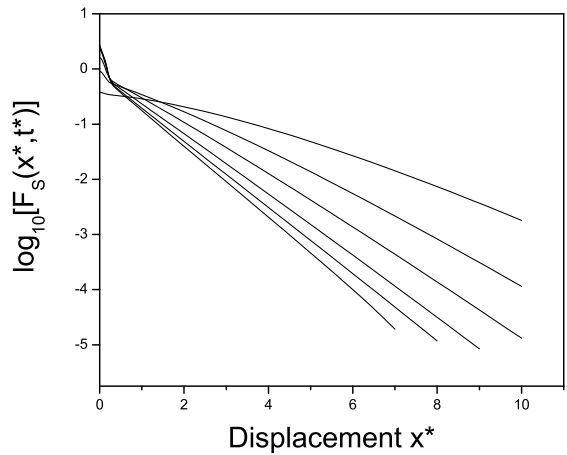}
\caption{\label{Gamma}  Weighted spatial distribution function $\bar F_s(x^*,t^*)$ as a function of the scaled displacement $x^*$ for scaled times $t^*= 0.03,\,0.1,\,0.3,\,1.0,\,3.0,\,10.0$, reading from left to right along the bottom of the graph.}
\end{figure}

Similarly,
\begin{equation}
\label{nMrt}
n_M(r^*,t^*)=- {1\over 4\,\pi\,r^*}\,{\partial\over\partial r^*}\, \Gamma_M(r^*,t^*),
\end{equation}
where 
\begin{equation}
\label{GammaM}
\Gamma_M(r^*,t^*)=\int_{-i\infty}^{+i\infty}{du\over 2\pi\,i u}\,e^{u t^*}\,\kappa(u)\,e^{-\,\kappa(u)\,r^*}.
\end{equation}
It is convenient to project these distributions onto, say, the $x^*$ axis, that is, to integrate out the perpendicular directions.  The formulas analogous to Eqs.(\ref{nGrt}) and (\ref{nMrt}) for the projected distributions, denoted $\bar n_G(x^*,t^*)$ and $\bar n_M(x^*,t^*)$, are
\begin{equation}
\label{nbarGxt}
\bar n_G(x^*,t^*) = \phi_G(t^*)\,\delta(x^*) + {1\over 2}\,\Gamma_G(x^*,t^*),
\end{equation}
and 
\begin{equation}
\bar n_M(x^*,t^*) = {1\over 2}\,\Gamma_M(x^*,t^*),
\end{equation}
where the functions $\Gamma_G$ and $\Gamma_M$ are the same as those given in Eqs.(\ref{GammaG}) and (\ref{GammaM}) with $r^*$ replaced by $x^*$.  

The $\delta$ functions in $n_G(r^*,t^*)$ and $\bar n_G(x^*,t^*)$ are the results of our having neglected the intra-cage motion discussed earlier.  In this approximation, with spatial resolution only of the order of $\ell$, molecules in frozen domains remain exactly at their initial positions with decaying probability $\phi_G(t^*)$.  The distributions $n_M(r^*,t^*)$ and $\bar n_M(x^*,t^*)$ have no $\delta$-function contributions because diffusion starts immediately in mobile regions. In graphs of experimental or computational data, these $\delta$ functions are visible as narrow Gaussian displacement distributions centered at $x^*=0$, whose root-mean-square widths, say $\ell_C^*=\ell_C/R^*$, are the average intra-cage  displacements that we are neglecting here.  

As a first look at the spatial distribution functions, in Fig.\ref{Gamma}, we show $\log_{10}[\bar F_s(x^*,t^*)]$ as a function of the displacement $x^*$ for a sequence of times $t^*$. In analogy to Eq.(\ref{ISFdef}), we write
\begin{equation}
\label{ISFxtdef}
\bar F_s(x^*,t^*)= {\cal P}_G\,\bar n_G(x^*,t^*)+ (1-{\cal P}_G)\,\bar n_M(x^*,t^*),
\end{equation}
For plotting these graphs, and for this purpose only, we replace the $\delta$ function in Eq.(\ref{nbarGxt}) by a normalized, Gaussian, intra-cage distribution
\begin{equation}
\label{barpC}
\delta(x^*) \to \bar p_C(x^*)= {1\over (2\,\pi\,{\ell_C^*}^2)^{1/2}}\,\exp\,\left(- {{x^*}^2\over 2\,{\ell_C^*}^2}\right).  
\end{equation}
We make this replacement primarily because the resulting graphs look -- and indeed are -- more realistic this way; but we emphasize that this is not a systematic correction of the small-time behavior because the intra-cage fluctuations are not otherwise included in $\bar n_G(x^*,t^*)$ or $\bar n_M(x^*,t^*)$.  Again, we choose $\Delta = 100$ and ${\cal P}_G=0.5$; and, so that the small-$x^*$, early-time behavior be visible in the figure, a relatively large but possibly realistic value of $\ell_C^*=0.1$.  Because of the factor $\phi_G(t^*)$ in Eq.(\ref{nbarGxt}), this central peak in $\bar n_G(x^*,t^*)$ disappears after times of the order of $\tau_{\alpha}^*$. 

At small times $t^*$, and for sufficiently large displacements $x^*$, the graphs in Fig.\ref{Gamma} exhibit the broad exponential tails reported elsewhere in the literature (see \cite{Weeks-Weitz-Science-00,CHAUDHURIetal07}).  The exponential tail is a robust mathematical feature of this class of models, closely associated with the decoupling of the mobile and glassy behaviors, and not dependent on any details of the glassy waiting-time distribution $\psi_G(t^*)$.  The degree of decoupling is reflected here by the magnitude of the parameter $\Delta$, which must be large for strong decoupling.  The only requirement on $\psi_G(t^*)$ is that it decays rapidly enough that $\tilde\psi_G(u)$ becomes vanishingly small at large $u$.

To see how the exponential function  emerges, evaluate the $u$-integrations in Eqs.(\ref{GammaG}) and (\ref{GammaM}) by first integrating around a circle of radius $u_0$ centered at the origin, and then closing the contour around the cut on the negative $u$-axis from $u=- u_0$ to $u \to -\infty$.  (The curves in Fig.\ref{Gamma} were computed numerically with $u_0 = 2$.) If $\Delta \gg 1$, we can choose $1 \ll u_0 \ll \Delta$, so that $\tilde\psi_G(u) \approx 0$ and $\kappa(u) \approx \sqrt{2}$ everywhere around the circle.  Then, for small $t^*$, the integration around the circle yields $\Gamma_G(x^*,t^*) \approx \Gamma_M(x^*,t^*) \approx \exp\,(- \sqrt{2}\,x^*)$.  This limiting behavior is most apparent in Fig.\ref{Gamma} for the two earliest times, $t^* = 0.03$ and $0.1$, where the slope has the predicted value of $-\sqrt{2}$. 

At larger times, and in the limit of indefinitely large $\Delta$, the preceding argument implies that all of these curves approach the same slope at large $x^*$.  On the other hand, at very early times $t^* \le \Delta^{-1}$, the integration is dominated by the discontinuity across the cut near $u = -\Delta$, and the displacement distribution is dominated by the early diffusive motion of the initially mobile molecules.  

Another interesting case is the limit of large $t^*$ and $x^*\ll t^*$.  The integrands in Eqs.(\ref{GammaG}) and (\ref{GammaM})  vanish like $\exp\,(- 1/w)$ in the limit $- u = w \to 0^+$.  Therefore, we can use the long-time, small-$w$ approximation, $A(w)\approx 1 + 3\,w/2$, to estimate
\begin{equation}
\label{nxlarget}
\Gamma_G(x^*,t^*)\approx \Gamma_M(x^*,t^*) \propto \exp\left(-{3\,{x^*}^2\over 4\,t^*}\right);
\end{equation}
thus the spatial  distribution reverts to a Gaussian.  Eq.(\ref{nxlarget}) fits the curve in Fig.\ref{Gamma} for $t^*=10$ reasonably well for $x^* < 5$.

\section{Moments of the Spatial Distributions}
\label{moments}

Another way of looking at these spatial distributions is to compute their time-dependent moments $\langle{r^*}^k(t^*)\rangle$. We use the identities
\begin{equation}
\langle {r^*}^2(t^*)\rangle = - 3\,\left[{\partial^2 F_s(k,t^*)\over \partial k^2}\right]_{k=0},
\end{equation}
and
\begin{equation}
\langle {r^*}^4(t^*)\rangle = 15\,\left[{\partial^4 F_s(k,t^*)\over \partial k^4}\right]_{k=0}.
\end{equation}

When inverting the Laplace transforms as in Eq.(\ref{hatndef}), we again close the contour of integration  in the negative $u$ plane, but in this case it is mathematically essential to include the circle around the origin in order to account for the singularities that occur there.  The mathematical situation becomes clear by inspection of the formula for $\langle {r^*}^2(t^*)\rangle_M$:
\begin{eqnarray}
\label{r2M}
\nonumber
\langle{r^*}^2(t^*)\rangle_M &=& 6\,\int_{- i\infty}^{+i\infty}{du\over 2\pi i}\,{e^{u\,t^*}\over \kappa^2(u)}\cr \\&=& 3\,\int_{- i\infty}^{+i\infty}{du\over 2\pi i}\,{e^{u\,t^*}\over 1 -\tilde \psi_G(u) + u/\Delta}.~~~~~
\end{eqnarray}
Because $\psi_G(u)\approx 1 - 3\,u/2$ for small $u$, this integrand has a first-order pole at the origin.  Higher-order moments have higher-order poles. 

\begin{figure}[h]
\centering \includegraphics[height=7 cm]{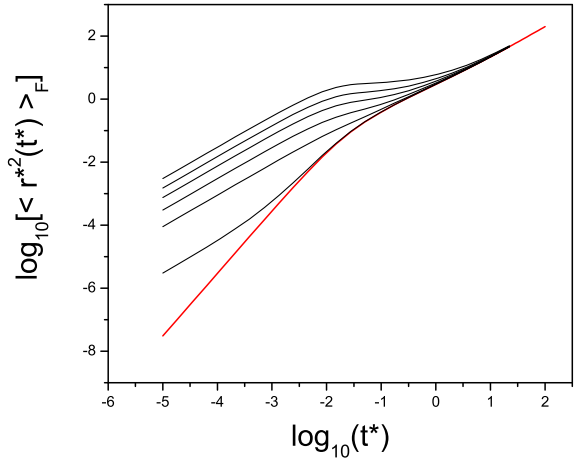}
\caption{\label{m2} (Color online) $\log_{10}[\langle{r^*}^2(t^*)\rangle_F]$ as a function of $\log_{10}(t^*)$ for ${\cal P}_G = 0,\,0.5,\,0.75,\,0.9,\,0.97,\, 0.999, {\rm and}\,\,1.0$ (red curve), reading from top to bottom.}
\end{figure}

We can use the analogs of Eq.(\ref{r2M}) for higher moments and for both the $n_G$ and $n_M$ distributions to obtain exact asymptotic results in the limits of  very small and very large $t^*$.  For vanishingly small values of $t^*$, close the contour around a circle at large values of $u$ where $\psi_G(u) \approx 2/u$. In this limit, the integrand behaves like $\exp\,(u\,t^*)/u^2$, and 
\begin{equation}
\label{m2M}
\langle{r^*}^2\rangle_M \approx 3\,\Delta\,t^*.
\end{equation}
Similar small-$t^*$ calculations yield:
\begin{equation}
\label{m4M}
\langle{r^*}^4\rangle_M \approx 15\,\Delta^2\,{t^*}^2;
\end{equation}
\begin{equation}
\label{m2G}
\langle{r^*}^2\rangle_G \approx 3\,\Delta\,{t^*}^2;
\end{equation}
and
\begin{equation}
\label{m4G}
\langle{r^*}^4\rangle_G\approx  10\,\Delta^2\,{t^*}^3.
\end{equation}

For very large $t^*$, close the contour in Eq.(\ref{r2M}) on a vanishingly small circle around the origin and use $\psi_G(u) \approx 1-3\,u/2$.  The result is that, after times so long that the tagged molecule no longer remembers whether it started in a mobile or a glassy region, both $\langle{r^*}^2\rangle_M$ and $\langle{r^*}^2\rangle_G$ converge to the same slowly diffusing Gaussian distribution for which $\langle{r^*}^2\rangle \approx 2\,t^*$  (with corrections of the order of $1/\Delta$). 

Note several features of these results.  As expected, the initially mobile molecules exhibit rapid (large $\Delta$) Gaussian diffusion at small times.  On the other hand, the displacements of the initially frozen molecules are non-Gaussian and ``pseudo-ballistic'' with $\langle{r^*}^2\rangle_G \sim \Delta\,{t^*}^2$. This behavior has nothing to do with early-stage ballistic motion of molecules within their cages but, rather, is an intrinsic feature of intermediate-stage, anomalous diffusion in this model.  

\begin{figure}[h]
\centering \includegraphics[height=7 cm]{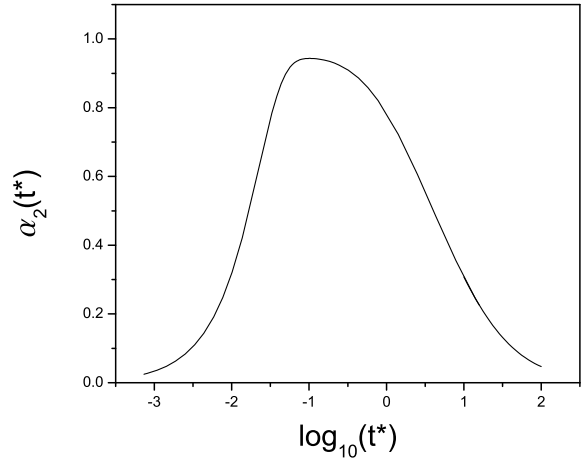}
\caption{\label{alpha2} Non-Gaussian parameter $\alpha_2$ as a function of $\log_{10}(t^*)$ for ${\cal P}_G =0$.}
\end{figure}
 
We turn finally to the full time dependence of the moments.  The results for the weighted average,
\begin{equation}
\langle{r^*}^2(t^*)\rangle_F = {\cal P}_G\, \langle{r^*}^2(t^*)\rangle_G + (1-{\cal P}_G)\,\langle{r^*}^2(t^*)\rangle_M,
\end{equation}
are shown in Fig.\ref{m2} for a series of different values of ${\cal P}_G$, as functions of $\log_{10}(t^*)$.  At very short times, these moments rise as functions of $t^*$ according to the  estimates in Eqs.(\ref{m2M}) - (\ref{m4G}).  The pseudo-ballistic behavior is apparent only for ${\cal P}_G \approx 1$ because, for smaller ${\cal P}_G$, these weighted moments are dominated by the displacements of the initially mobile molecules.  After the initial rise, at roughly $t^*\approx \Delta^{-1} = 10^{-2}$, the moments cross over from fast $\beta$ relaxation to slow $\alpha$ decay as the tagged molecules are repeatedly trapped and then escape from glassy domains.  Finally, in Fig.\ref{alpha2}, we show the non-Gaussian parameter
\begin{equation}
\alpha_2(t^*) = {3\,\langle{r^*}^4(t^*)\rangle_M\over5\,\langle{r^*}^2(t^*)\rangle_M^2}-1 
\end{equation}
only for ${\cal P}_G = 0$ because, in this approximation where we have neglected intra-cage vibrational motions, only the mobile molecules exhibit Gaussian displacement distributions at early times.  Here we see explicitly that the non-Gaussian behavior occurs during the crossover from $\beta$ to $\alpha$ relaxation.  Both of these last two figures are artificial in the sense that we are independently varying the parameter ${\cal P}_G$, which in fact ought to be determined uniquely by the temperature along with the parameters $R^*$ and $\Delta$.  

\section{Concluding Remarks}

The model proposed here contains a relatively simple mechanism for producing stretched exponential decay of molecular correlations, as observed via the self intermediate scattering function. That mechanism emerges directly from the spatial heterogeneity of a glass-forming liquid.  The model illustrates how anomalous diffusion, as exemplified by a broad exponential (non-Gaussian) tail of the molecular displacement distribution, is related to heterogeneity and -- indirectly -- to the accompanying stretched exponential behavior.  Yet another feature of the model is the way in which it illustrates how the transition from relatively fast $\beta$ relaxation to slow $\alpha$ decay correlates with the onset of anomalous diffusion.  Our analysis neglects the very short time, strongly localized, intra-cage fluctuations.  The transition between the latter motions and cage-breaking events seems to us to be outside the range of validity of the continuous-time random-walk theory used here.  

There are, of course, many missing ingredients, almost all of which pertain to the temperature dependence of the various parameters introduced here.  One of us (JSL) plans to address those issues in a subsequent report.

\begin{acknowledgments}
This research was supported by U.S. Department of Energy Grant No. DE-FG03-99ER45762. 
\end{acknowledgments}

\end{document}